# Primordial black holes and the Asymmetrical Distribution of Short GRB Events*

David B. Cline

*Department of Physics and Astronomy, Box 951547*
*University of California Los Angeles*
*Los Angeles, CA 90095-1547, USA*

**Abstract**. We review some of the expectations for Primordial Black Hole evaporation and we review the properties of short Gamma Ray Bursts. It is plausible that these GRB come from PBH evaporation. We then show that the angular distribution asymmetry recently observed by us could be due to a concentration of PBH in the Galactic Arms like the Orion arm.

## 1. Introduction

The search for evidence for primordial black holes (PBHs) has continued since the first discussion by Hawking [1]. In fact, this was about the time that gamma-ray bursts (GRBs) were first discovered, making a natural association with PBHs. However, in the intervening years it has become clear that the time history of the typical GRB is not consistent with the expectations of PBH evaporation [2].

While the theory of PBH evaporation has been refined, there are still no exact predictions of the GRB spectrum, time history, etc. However, reasonable phenomenological models have been made, and the results again indicate that most GRBs could not come from PBHs [3]. In addition, there are new constraints on the production of PBHs in the Early Universe that indicate that the density of PBHs in the Universe should be very small, but not necessarily zero [4].

After the initial discovery of GRBs, it took many years to uncover the general properties. Around 1984, several GRBs were detected, indicating that there was a class of short bursts with dime duration of ~ 100 ms and a very short rise time [5]. A separate class of GRBs was declared. This classification seems to have been forgotten and then rediscovered by some of us.

It is still possible that there is a sizable density of PBHs in our Galaxy and that some of the GRBs could be due to PBH evaporation. Recently we showed that the BATSE 1B data have a few events that are consistent with some expectation of PBH evaporation (short bursts with time duration <200 ms and that are consistent with $V/V_{max}$ ~ ½ .

## 2. Primordial Black Hole Evaporation and Fireball

Ever since the theoretical discovery of the quantum-gravitational particle emissions from black holes by Hawking, there have been many experimental searches (see Ref. [ 6 ] for details) for high-energy gamma-ray radiation from PBHs. They would have been formed in the Early Universe and would now be entering their final stages of extinction. The violent final-stage evaporation or explosion is the striking result of the expectation that the PBH temperature is inversely proportional to the PBH mass, e.g., $T_{PBH} \approx 100$ MeV $(10^{15}$ g$/m_{PBH})$, since the black hole becomes hotter as it radiates more particles and can eventually attain extremely high temperatures.

---

In 1974, S. Hawking showed in a seminal paper that an uncharged, non-rotating black hole emits particles with energy between $E$ and $E + dE$ at a rate per spin of helicity state of

$$\frac{d^2N}{dtdE} = \frac{\Gamma_s}{2\pi h} \left[ \exp\left(\frac{8\pi GME}{hc^3}\right) - (-1)^{2s} \right]^{-1}, \quad (1)$$

where $M$ is the PBH mass, $s$ is the particle spin, and $\Gamma_s$ is the absorption probability [1]. It can be considered that this particle emission comes from the spontaneous creation of particle–antiparticle pair escapes to infinity, while the other returns to the black hole. Thus, the PBH emits massless particles, photons, and light neutrinos, as if it were a hot black-body radiator with temperature $T \cong 10^{16}$ g/$M$ MeV, where $M$ is the black hole mass [3]. A black hole with one solar mass, $M_1 \cong 2 \times 10^{33}$ g, has an approximate temperature of $10^{-3}$ K, while a black hole with mass of $6 \times 10^{14}$ g has a temperature of $\sim 20$ MeV. The temperature of a black hole increases as it loses mass during its lifetime. The loss of mass from a black hole occurs at a rate, in the context of the standard model of particle physics, of

$$\frac{dM}{dt} = -\frac{\alpha(M)}{M^2}, \quad (2)$$

where $\alpha(M)$, the *running constant*, counts the particle degree of freedom in the PBH evaporation. The value of $\alpha(M)$ is model dependent. In the standard model (SM), with a family of three quarks and three leptons, it is given [2,6,13] as $\alpha(M) = (0.045\, S_{j=+1/2} + 0.162\, S_{j=1}) \times 10^{-4}$, where $S_{j=+1/2}$ and $S_{j=1}$ are the spin and color degrees of freedom for the fermions and gauge particles, respectively. For the SM, $\alpha(M) = 4.1 \times 10^{-3}$.

A reasonable model of the running coupling constant is illustrated in Figure 1, where the regions of uncertainty are indicated [2][7]. These are the regions where there could be a rapid increase in the effective number of degrees of freedom due to the quark–gluon phase (QGP) transition. The phase transition would lead to a rapid burst in the PBH evaporation or, at high energy, there could be many new particle types that would also lead to an increase in the rate of evaporation. Also shown in Figure 1 are the regions in PBH temperature where short duration GRBs may occur when the PBH mass is either $10^{14}$ or $10^9$ g.

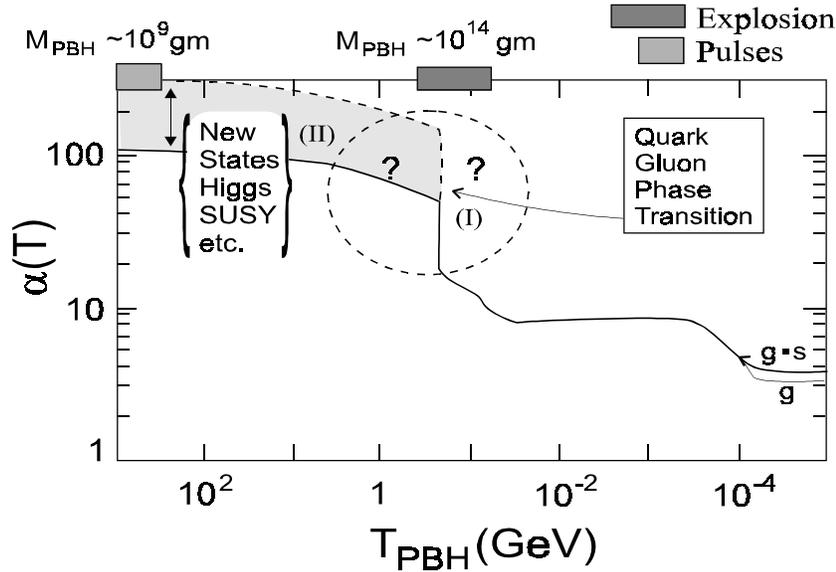

Figure 1. Running coupling or density of states factor $\alpha$, showing regions of uncertainty due to the QGP transitions (I) or the increase in the number of new elementary particles (II). It is possible that intense short GRBs could occur at either of these temperatures (the decrease of mass of the black hole is given by $dm_{PBH}/dt = -\alpha/m_{PBH}^2$). A rapid mass decrease or burst can occur when $m_{PBH} \leq 10^{10}$ g or if $\alpha$ changes rapidly near the QGP transition.

Black holes at the evaporation stage at the present epoch can be calculated as having $M_* \cong [3\alpha(M_*)\tau_{evap}]^{1/3} \cong 7 \times 10^{14}$ g for $\alpha(M_*) \cong 1.4 \times 10^{-3}$. The bound on the number of black holes at their critical mass, constrained by the observed diffuse gamma-ray background, has been put in the 10–100 MeV energy region [4]:

$$N = \left.\frac{dn}{d(\ln M)}\right|_{M=M_*} \leq 10^5 \, \text{pc}^{-3} \ . \tag{3}$$

Thus, the number of black holes with critical mass $M_*$ in their final state of evaporation is

$$\frac{dn}{dt} = -\frac{3\alpha(M_*)}{M_*^3} = N = 2.2 \times 10^{-10} \, N \, \text{pc}^{-3} \, \text{yr}^{-1} \ . \tag{4}$$

Based on previous calculations and numerous direct observational searches for high-energy radiation from an evaporating PBH, we might conclude that it is not likely to single be able to single out such a monumental event. However, we pointed out a possible connection between very short GRBs observed by the BATSE team and PBH evaporation emitting very short energetic gamma rays [2][6]. If we want to accept this possibility, we may have to modify the method of calculating the particle emission spectra from an evaporating PBH, in particular, at or near the quark–gluon plasma], for a review, and phase transition temperature at which the $T_{PBH}$ arrives eventually. We briefly discussed that inclusion of the QGP effect around the evaporating PBH at the critical temperature may drastically change the resulting gamma-ray spectrum [7][6]. The QGP interactions around the evaporating PBH form an expanding hadronic (mostly pions) matter fireball. Shortly after the decay of pions, the initial hadronic fireball converts to a fireball with mixtures of photons, leptons, and baryons.

Using the simplest picture, i.e., only $\pi^0$s produced in the QGP transition, we can obtain the properties of the fireball. We find out that given that $L_{PBH} \sim L_{QGP} \sim 5 \times 10^{34}$ ergs and $T_{PBH} \sim T_{QGP} \geq 160$ MeV, a simple radiation-dominated model would give $\tau_s \sim 10^9$ cm, which implies that $\tau$ is of order 100 ms. However, this is very uncertain and could even be the order of seconds in some cases. Thus one can expect GRBs from a fireball to have both a very short rise time ($\leq 1$ µs) and duration ($\sim$50–200 ms) also in this model.

## 3. Study of Short GRBs [8] *

We have studied all events with $T_{90}$ less than 200 ms and then refit the time profile using the TTE data with the BATSE 3B data set (we restrict this part of the analysis to BATSE 3B data. We found 12 events that have a fitted time duration of less than 100 ms. These events are listed in Table 1 with the fitted time duration. We then closely inspected the 12 events and found that some have additional structure. Of the 10 good/fair events, 8 have a time duration of 66 ms or less. Our goal is to select a similar class of events, so we only study single-peak events which constitute the bulk of short bursts. As we will show, these events appear to be almost identical in all features, expect for BATSE trigger 2463, which we delete in this analysis. This event has a clearly different energy spectrum from the bulk of the short GRB events. This sample constitutes the set of events studied here.

___________________

* This section adapted from reference 8.

# TABLE 1
## EVENT SELECTION AND PROPERTIES (BATSE 3B)

| Trigger Number | Duration from TTE Fit (s) | Hardness Ratio | Comments | Used in This Analysis |
|---|---|---|---|---|
| 01453...... | 0.006 ± 0.0002 | 6.68 ± 0.33 | Poor event, precursor | No |
| 00512...... | 0.014 ± 0.0006 | 6.07 ± 1.34 | Good event | Yes |
| 01649...... | 0.020 ± 0.0080 | .................. | Fair event* | Yes |
| 00207...... | 0.030 ± 0.0019 | 6.88 ± 1.93 | Good event | |
| 02615...... | 0.034 ± 0.0032 | 5.42 ± 1.15 | Good event | Yes |
| 03173...... | 0.041 ± 0.0020 | 5.35 ± 0.27 | Poor event, precursor | No |
| 02463...... | 0.049 ± 0.0045 | 1.60 ± 1.55 | Good event | No |
| 00432...... | 0.050 ± 0.0018 | 7.46 ± 1.17 | Good event | Yes |
| 00480...... | 0.062 ± 0.0020 | 7.14 ± 0.96 | Good event | Yes |
| 03037...... | 0.066 ± 0.0072 | 4.81 ± 0.98 | Good event | Yes |
| 02132...... | 0.090 ± 0.0081 | 3.64 ± 0.66 | Good event | Yes |
| 00799...... | 0.097 ± 0.0101 | 2.47 ± 0.39 | Fair event* | No |

* Fair event, small additive structure

     To obtain a better understanding of the short bursts, we discuss two individual bursts that have more detailed information than the bulk of the events. According to the arguments given above, we expect all the short burst to be very similar and, therefore, we assume that the behaviour of these special bursts is likely an example for all short bursts. If these events are typical of the short bursts, then we can see a clear behaviour in the fine time structure and the detected $\gamma$ energy distribution. We start with the incredible GRB trigger 512. We note the detailed fine structure for BATSE trigger 512, which has the finest time structure of any GRB observed to date -- possibly down to 20 $\mu s$ level. This was a very bright burst and allowed unpredicted time information.

     The PHEBUS GRB detection (see PHEBUS Catalog in Terekov et al. 1994 [8]) has recorded two very interesting short time events shown in Figure 2. As far as can be determined, these events are identical (note that the energy distribution are fit to a synchrotron and are identical). Because of a thicker absorber, the PHEBUS detector has a larger energy-range capacity than that of the BATSE detector.

     Thus, this detector can record photon energies up to 180 MeV in contrast to BATSE, which is only sensitive to about 580 keV due to the absorbers. Note that these two PHEBUS events have photon energies above 1 MeV. Thus, the short GRBs have energetic photons in the spectrum.

     We classify all GRBs into three different categories: one with $\tau > 1s$ (long, L), one with $1s > \tau > 0.1s$ (medium, M), and one with $\tau \leq 100$ ms (short, S), which is the focus of this investigation.

     We note that the short bursts are strongly consistent with a $C_p^- 3/2$ spectrum, indicating a Euclidean source distribution, as was shown previously in reference 6. In the medium (100 ms to 1 s) time duration, the ln $N$ - ln $S$ distribution seems to be non-Euclidean; in the long duration ($\tau > 1$ s) bursts, the situation is more complicated as we have shown recently [8]. The $<V/V_{max}>$ for the S, M, and L class of events is, respectively, 0.52 ± 0.1, 0.36 ± 0.02, 0.31 ± 0.01. (In this case, we have used the BATSE 5B to obtain the best statistics.)

     We have shown that the GRBs with $\tau < 100$ ms likely are due to a separate class of sources and appear to be nearly identical in contrast to the bulk of GRBs. In this analysis, we have studied a small class of BATSE 3B TTE data in detail, and to improve the statistical power for some issues, we have used BATSE 4B and the latest BATSE 5B data. We do not believe this study warrants the use of the full TTE data for BATSE 4B or BATSE 5B, since the point is to show a general morphology of the GRBs, not a complete statistical analysis at this stage. It is likely that the source is

local or Galactic, in contrast to the cosmological origin of the bulk of GRBs. One model source that may produce such a unique class of GRBs is the evaporation of PBHs. Independent of that model, we believe these short bursts constitute a third class of GRBs.

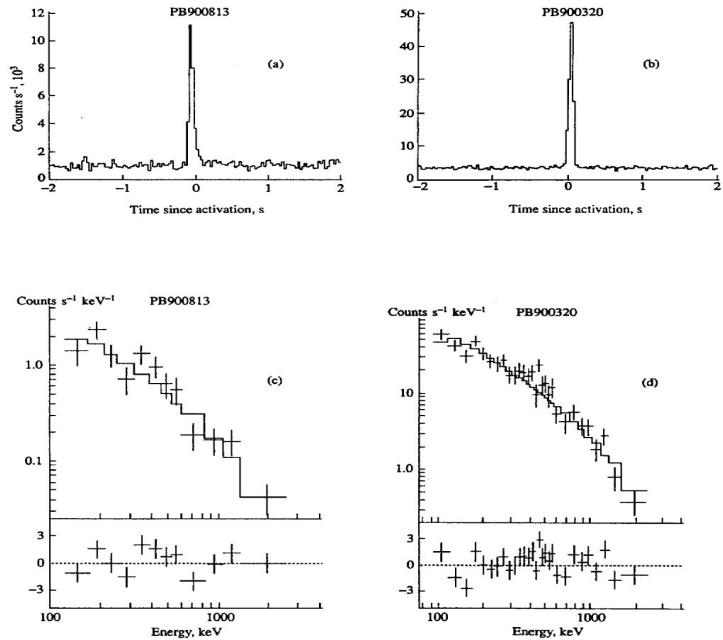

Figure 2. Two short bursts from the PHEBUS detector: Time profiles of (a) PB 900813 and (b) PB 900320, and energy spectrum of (c) PB 900813 and (d) PB 900320 [8].

## 4. Possible Rate for GRB from PBH Evaporation

Over the past two decades, the reality of a diffuse component of the gamma-ray flux has been established. While there is no firm explanation of the source of these gamma rays, one possible candidate is due to the evaporation of PBHs in the Universe. In fact, the diffuse gamma-ray spectrum and flux have been used to put the only real limit on the density of PBHs in the Universe, leading to a limit of $\Omega_{PBH} < 10^{-7}$. The recent work of Dixon et al [9] has claimed the existence of an important component of gamma rays in the Galactic halo. This method of analysis is very different from previous analyses and gives some confidence that the results are consistent with those of other experiments. This suggests that there is a halo component of diffuse gamma rays. The flux level of this component is very similar to the extragalactic diffuse flux.

We now turn to a simple model that explains the current observations by assuming the existence of PBHs at the level of the relaxed Page–Hawking bound discussed previously and the Galactic clustering enhancement factors of $5 \times 10^5$ or greater. In Figure 3, we show the logic of the possible detection of individual PBHs by very short GRBs and the connection with the diffuse gamma-ray background. The rate is consistent with a few PBH evaporation for $(Pc)^3$ in our galaxy.

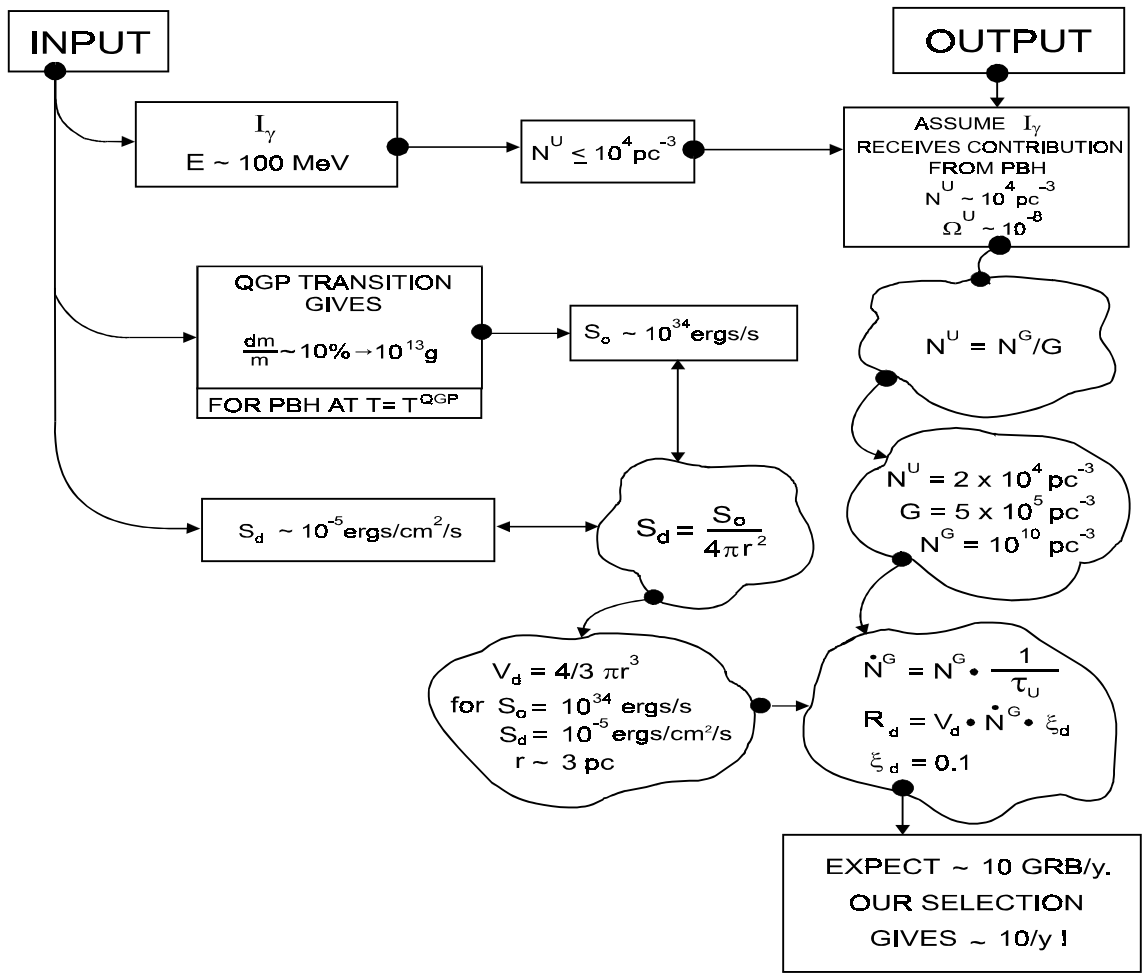

Figure 3. Schematic of the connection between GRB events and the diffuse γ background, $I_\gamma$, where $N^U$ is density of PBHs in the Universe, $\Omega^U$ is ratio of density of the Universe, $T^{QGP}$ is QGP transition temperature, $S_o$ is GRB luminosity from PBH, $N^G$ is number of PBHs in the Galaxy, G is galaxy clumping factor, $S_d$ is fluence sensitivity of GRB detectors, r is distance to the source, $\dot{N}^G$ is rate of PBH decay in the Galaxy, $\tau_U$ is age of the Universe, $R_d$ is number of GRBs detected per year from PBHs, $\xi_d$ is GRB detector efficiency (including the fraction of energy detected).

## 5. Discovery of an Angular Asymmetry of Short GRBs **

Figure 4 shows the time distribution T90 for all GRBs from the BATSE detector up to Nov. 1998. Hereafter we will use this data. We divide the GRBs into three classes in time duration: long, L ($\tau > 1$ s); short, M (1 s $> \tau > 0.1$ s); and very short, S ($\tau < 100$ ms). We use the duration time of T90 for all of this analysis. Henceforth in this letter, we confine the discussion to the M and S classes of GRBs.

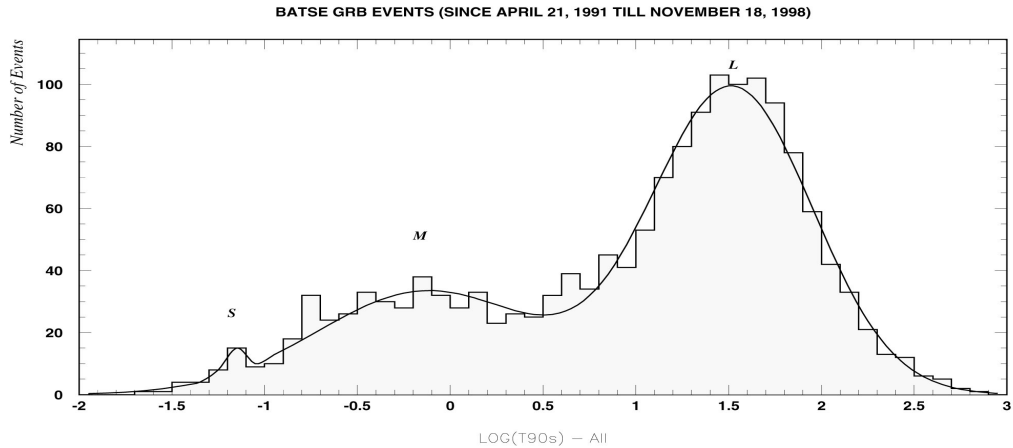

Figure 4. Time distribution T90 for all GRB events. The line is a result of three Gaussian curve fit.

Since these events are adjacent in time, it is important to contrast the behaviour.

We note that the short bursts are strongly consistent with a $\bar{C_p} 3/2$ spectrum, indicating a Euclidean source distribution, as was shown recently by ( [8]). In the medium (from 100 ms to 1 s) time duration, the ln N - 1 - S distribution seems to be non-Euclidean; in the long duration ($\tau > 1$ s) bursts, the situation is more complicated as we have shown recently ( [8]). The $<V/V_{max}>$ for the S, M, and L class of events is, respectively, 0.52 ± 0.06, 0.36 ± 0.02, 0.31 ± 0.01.

One possibility for explaining these effects is that a big part of the short bursts may come from a local Galactic source. This explanation is likely not viable for the medium time bursts, since $<V/V_{max}>$ is 0.36 ± 0.02, indicating a likely cosmological source. The longer bursts are clearly mainly from cosmological distance.

We assume that the S GRBs constitute a separate class of GRBs and fit the time distribution in Figure 4 with a three-population model. The fit is excellent but does not in itself give significant evidence for a three-population model.

We now turn to the angular distributions of the S and M GRBs. In Figure 5A we show this distribution for the very short bursts. We can see directly that this is not an isotropic distribution. To ascertain the significance of the anisotropy, we break up the Galactic map into eight equal probability regions. In Figure 2B, we show the distribution of events in the eight bins; clearly one bin has a large excess. To determine the statistical probability for such a deviation, we calculate the Poisson probability distribution for the eight bins with a total of 42 events. This distribution is also shown in Figure 2B.

---
* Adapted from a UCLA report by D. Cline, C. Matthey, So. Otwinowski

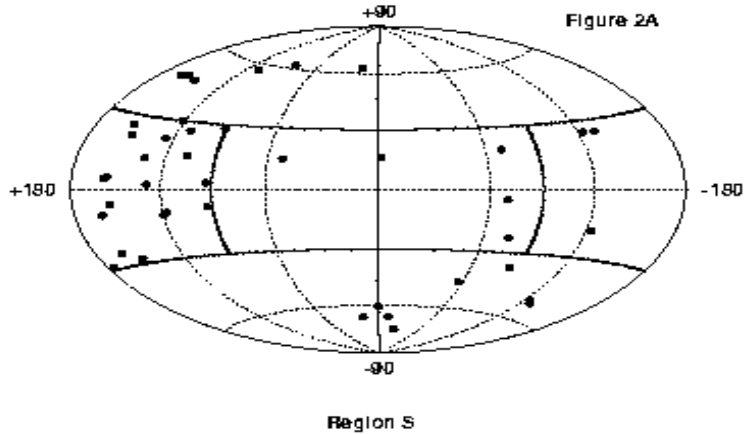

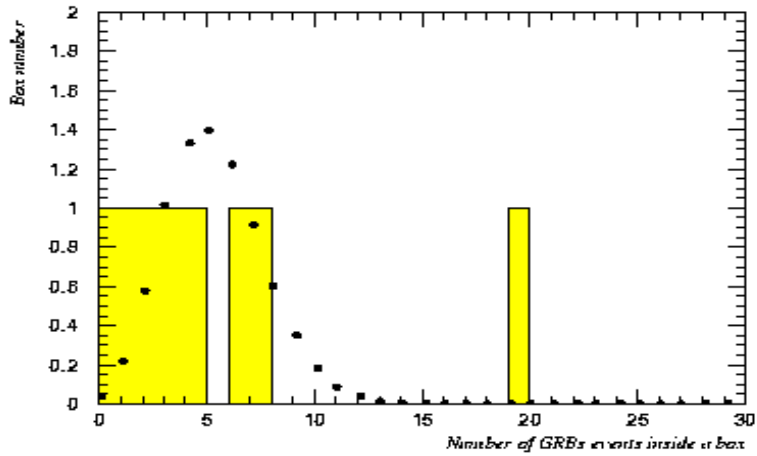

Figure 5 (A). The Galactic coordinate angular distribution of the GRB events with very short time duration (S). We define eight regions that correspond to equal angular surface. (B) The distribution of the number of GRB events in each of the eight regions. Points correspond to prediction of Poisson distribution for 42 events divided into eight equiprobable groups. Both distributions are normalized to the same surface.

The probability of observing 19 events in a single bin is $1.6\times10^{-5}$; we neglect the non-uniformity in sky exposure, nevertheless, we consider this a very significant deviation from an isotropic distribution. We note that no cuts were made specifically to search for a non isotropic distribution. The sample was selected for other reasons based on the study of short duration GRB's.

To contrast the distribution of the S GRBs and to test for possible errors in the analysis, we plot the same distributions for the M sample in Figures 6A and 6B. As can be seen from Fig. 6A, this distribution is consistent with isotropy. Figure 6B shows the same analysis as Figure 5B, indicating that there are no bins with a statistically significant deviation from the hypothesis of an isotropic distribution. Of the 42 very short S GRBs, there are 19 in the excess region (Figure 5)..

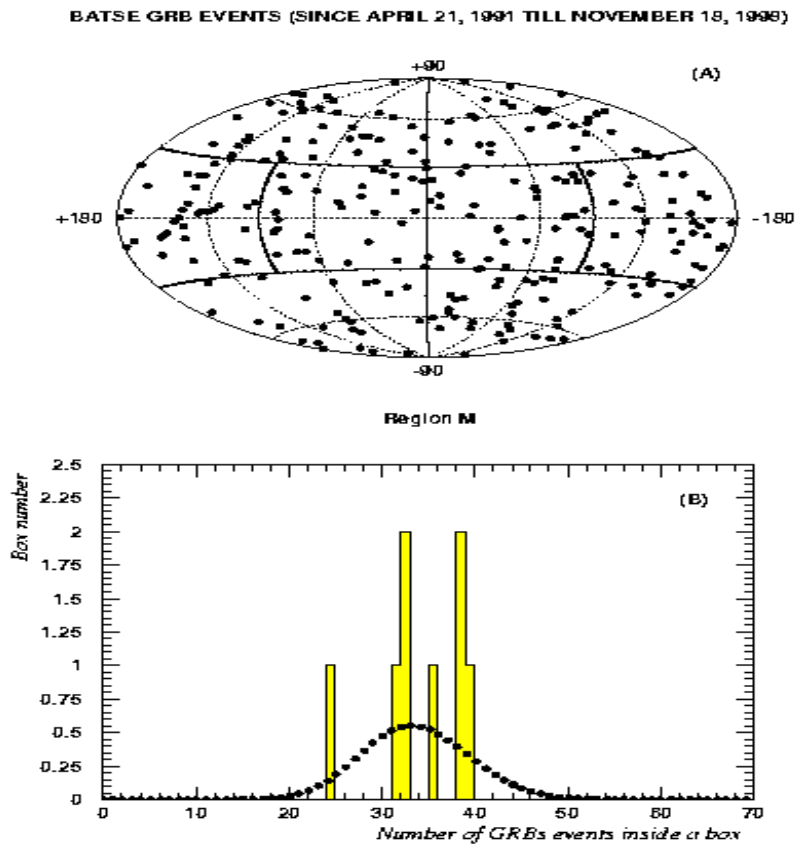

Figure 6. The Galactic coordinate angular distribution of the GRB events with short time duration (M). Eight regions correspond to equal angular surface (like those in Figure 2A). (B) The distribution of the number of GRB events in each of the eight regions. Points correspond to prediction in the Poisson (Gauss) distribution for 269 events divided into eight equiprobable groups. Both distributions are normalized to the same surface.

## 6. Possible Association with the Orion Arm of the Galaxy [10]

We have presented evidence in this report that the very short time GRBs form a separate class of events - of non-cosmological origin. Note that the rest of the GRBs (class M and L) are fully consistent with a cosmological origin.

In table 2 we summarize this evidence. In addition the angular distribution asymmetry could be explained if an excess of sources were in the Orion Galaxy arm as shown in Fig. 7. It is possible that Primordial Black Holes could be concentrated in these Galactic arms. However, the rest of the events are isotropic, and are thus consistent with a uniform distribution of PBH in the Galaxy.

We thank Stan Otwinowski and Christina Matthey for their help with this work.

_________________________________________________________________________________

**TABLE 2**
**Arguments for a New Exotic Class of GRBs in the Short Bursts**
_________________________________________________________________________________

i) Time variations in some events (trigger 512) of ~ (10-20) $\mu$s suggest the size of the source is

$$\Delta \gamma \leq \Delta t \times t \leq 6 km$$

(much smaller than any astrophysical source except Black Holes) [8]

ii) The *ln N - ln S Distribution* suggests a local source ($<V/V_{max}> = 0.52 \pm 0.00$) [8]

iii) The *Angular Distribution* not consistent with a cosmologcal origin – and could be consistent with an enhancement of sources in the Orion Galactic Arm [10]

iv) The *Nearly Identical Nature* of the event suggests an exotic source [8]

_________________________________________________________________________________

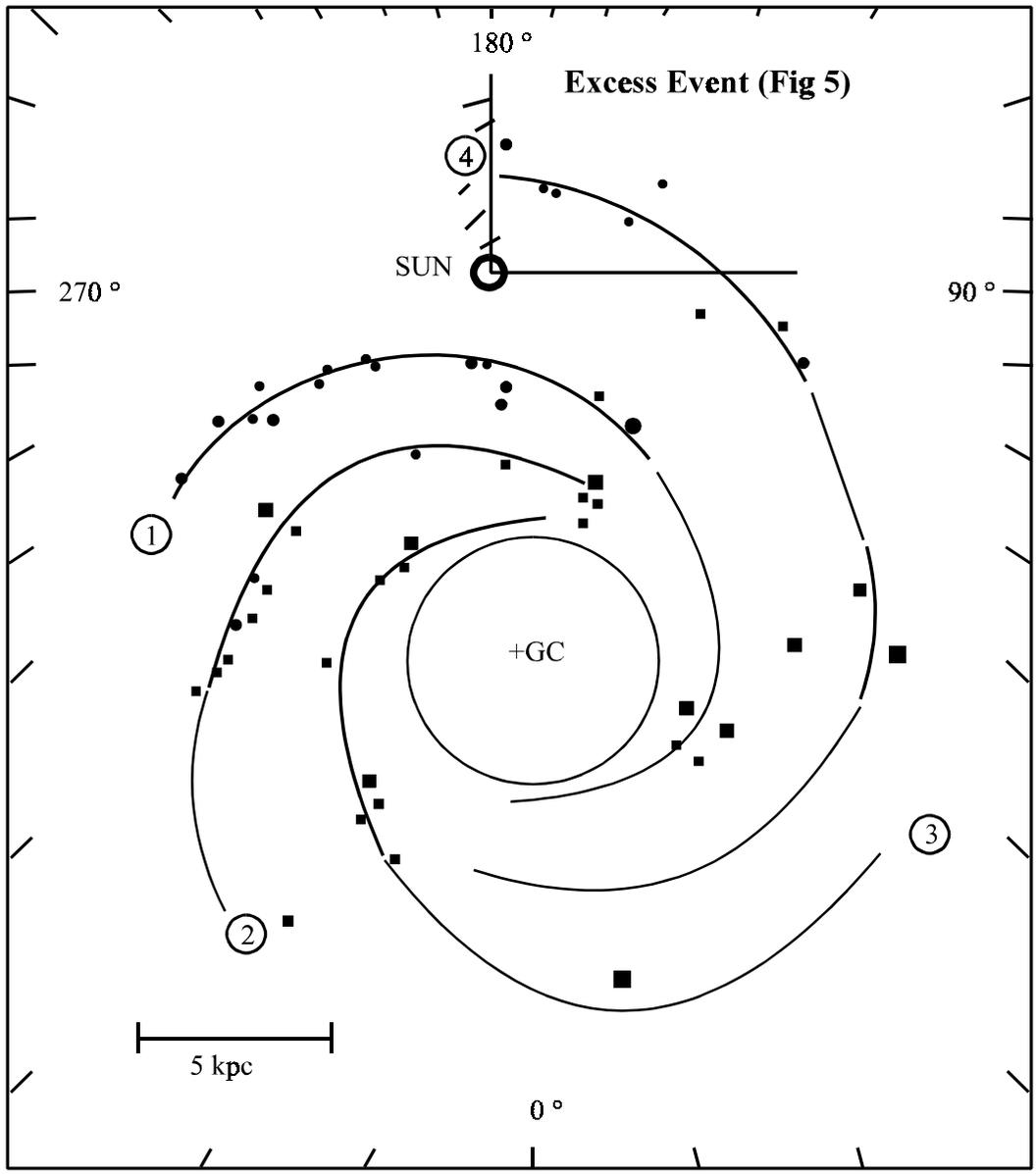

Figure 7. The excess events could come from one of the Galactic Arms.

# REFERENCES

1. Hawking, S. W. 1974, Nature, 30, 248.

2. Cline, D. B. and Hong, W. P. 1992, ApJ, 401, L57.

3. Page, D. N. 1976, Phys. Rev. D, 13, 198.

4. MacGibbon, J. H., & Carr, B. J. 1991, ApJ, 371, 447.

5. Barat, C., Hayles, R. I., Hurley, K., Niel, M., Vedrenne, G., Estulin, I. V., & Zenchenko, M. 1984, ApJ, 285, 791.

6. Cline, D. B., Matthey, C., & Otwinowski, S. 1998b, in Proc. 14th IAP Colloq. (Paris: Editions Frontieres), 374.

7. Cline, D. B. 1996, Nucl. Phys. A, 610, 500.

8. Cline, D. B., Matthey, C., & Otwinowski, S. 1999, ApJ, 527, 825.

9. Dixon, D. 1997, quoted in "Halo Around Milky Way Is Reported" by Cole, K. C. Nov. 5, Los Angeles Times, 116 A1.

10. Report in Progress, D. Cline, C. Matthey & S. Otwinowski.